\documentclass[structabstract]{aa}  
\usepackage{graphicx}
\usepackage{txfonts}
\usepackage{longtable}
\usepackage[authoryear]{natbib}
\bibpunct[; ]{(}{)}{;}{a}{}{,}

\begin{document}
   \title{Orbital and physical parameters of eclipsing binaries from the ASAS catalogue}
   
   \subtitle{II. Two spotted $M<1$ M$_\odot$ systems at different evolutionary stages.}

   \titlerunning{Two spotted $M<1$ M$_\odot$ systems from ASAS}

   \author{K. G. He\l miniak\inst{1}
          \and
          M. Konacki\inst{1,2}
          }

   \institute{Nicolaus Copernicus Astronomical Center, Department of Astrophysics,
              ul. Rabia\'nska 8, 87-100 Toru\'n, Poland\\
              \email{xysiek@ncac.torun.pl, maciej@ncac.torun.pl}
         \and
             Astronomical Observatory, Adam Mickiewicz University, ul. S\l oneczna 36,
		60-286 Pozna\'n, Poland\\
             }

   \date{Received ...; accepted ...}

 
  \abstract
   {}
   {We present the results of our detailed spectroscopic and photometric analysis
of two previously unknown $< 1$ M$_\odot$ detached eclipsing binaries: 
\object{ASAS J045304-0700.4} and \object{ASAS J082552-1622.8}.}
   {With the \emph{HIgh Resolution Echelle Spectrometer} (HIRES) on the Keck-I 
telescope, we obtained spectra of both objects covering large fractions of orbits
of the systems. We also obtained $V$ and $I$ band photometry with the 1.0-m 
Elizabeth telescope of the South African Astronomical Observatory (SAAO).
The orbital and physical parameters of the systems were derived with the PHOEBE and 
JKTEBOP codes. We investigated the evolutionary status of both binaries
with several sets of widely-used isochrones.}
   {Our modelling indicates that (1) \object{ASAS J045304-0700.4} 
is an old, metal-poor, active system with component masses of 
$M_1 = 0.8452 \pm 0.0056$ M$_{\odot}$, 
$M_2 = 0.8390 \pm 0.0056$ M$_{\odot}$ and radii of
$R_1 = 0.848 \pm 0.005$ R$_{\odot}$ and 
$R_2 = 0.833 \pm 0.005$ R$_{\odot}$, which places it at the end of the 
Main Sequence evolution -- a stage rarely observed for this type of stars. 
(2) \object{ASAS J082552-1622.8} is a metal-rich, 
active binary with component masses of 
$M_1 = 0.703 \pm 0.003$ M$_{\odot}$, 
$M_2 = 0.687 \pm 0.003$ M$_{\odot}$ and radii of
$R_1 = 0.694 ^{+0.007}_{-0.011}$ R$_{\odot}$ and
$R_2 = 0.699 ^{+0.011}_{-0.014}$ R$_{\odot}$. 
Both systems show significant out-of-eclipse variations, 
probably owing to large, cold spots. we also investigated the influence of 
a third light in the second system.}
   {}


   \maketitle
%

\section{Introduction}

Eclipsing binaries provide the absolute values for a number of physical 
parameters of stars that are crucial for testing the stellar structure 
and for evolution models. These parameters are especially needed for stars 
with masses below 1 M$_\odot$. They are the most numerous in the Galaxy, 
yet our models often fail to accurately reproduce their properties 
\citep{rib08}. In order to perform reliable tests of the models, 
we need to derive masses, radii, and other parameters with a precision 
of 3\% or better \citep{bla08,cla08}. In recent years progress 
in the detection and characterization of the K and M-type eclipsing 
binaries was made \citep{shk08,cak09a,irw09}. But not every 
binary from the limited sample of low-mass systems satisfies the 
3\% accuracy criterion, which makes every new datapoint important.

In this paper we present for the first time the orbital and physical 
parameters of two new detached eclipsing spectroscopic binaries (DEBs), 
both of which have the component masses below 1 $M_{\odot}$ -- 
\object{ASAS J045304-0700.4} (hereafter ASAS-04) and 
\object{ASAS J082552-1622.8} (hereafter ASAS-08). We reach 
0.5 \% level of precision for the masses and radii. 
Both objects seem to be interesting from the evolutionary point 
of view. ASAS-04 is most likely an active, old, metal-poor and spotted 
object that seems to be near the end of its main-sequence stage. 
ASAS-08 on the contrary is younger, metal rich, more active 
and covered with larger spotted areas. we present our observations 
and their analysis below. 


\section{ASAS-04 and ASAS-08}

\subsection{ASAS-04 (\object{GSC 04749-00560})}
The eclipsing binary \object{ASAS J045304-0700.4} is the fainter
of the two objects -- its apparent $V$ magnitude depending on the 
catalogue is between 11.125 \citep{kha01} and $11.35$ mag \citep{poj02}. 
Our photometry (Sect. \ref{sect_phot}) indicates $V = 11.240\pm0.005$ 
and $I= 10.374\pm0.005$ mag at the maximum. Its eclipsing nature was 
noted for the first time in the \emph{ASAS Catalog of Variable 
Stars}\footnote{http://www.astrouw.edu.pl/asas/} 
\citep[ACVS;][]{poj02}. The system is classified in ACVS as an 
eclipsing detached  binary (ED) with an amplitude of the photometric 
variation of $0.29$ mag. The ASAS photometric dataset now spans more than 
nine years and provides a complete phase coverage. A quick look at the 
ASAS light curve allows one to conclude that the given period 
$P = 1.6224\,d$ can be improved. When phased with the correct 
period (see Table \ref{tab_orb}), the ASAS curve shows no clear 
out-of-eclipse variations. ASAS-04 is likely associated with the 
faint ROSAT X-ray source \object{1RXS~J045304.1-070011}
($F_x = 1.073 \times 10^{-12}$ erg cm$^{-2}$ s$^{-1}$).
The total ROSAT positional error of 12'' for this source 
is a bit smaller than the separation between ROSAT and the ASAS-04 
position in the visual ($\sim 13$'', even including a small but 
measurable proper motion).

\subsection{ASAS-08 (\object{GSC 05998-01918}, \object{EUVE J0825-16.3})
	\label{sec_lit}}
This system's magnitude in $V$ available from various sources is between 10.20 
\citep{poj02} and $10.42$ mag \citep{urb01}. Our photometry -- 
$V = 10.142\pm0.003$ and $I= 8.708\pm0.003$ mag at the maximum -- seems 
to favour the first value. 
The binary was identified as a faint extreme ultraviolet \citep{lam97}
and a soft X-ray source \citep[\object{1RXS~J082551.4-162244},
$F_x = 3.556 \times 10^{-12}$ erg cm$^{-2}$ s$^{-1}$; ][]{vog99}. It was later
optically identified as a dKe or dMe star \citep{pol97}. ASAS-08 was  
reported as a double-lined spectroscopic binary (SB2) by \citet{chr02}, 
who also provided the equivalent widths of the $H_{\alpha}$ and $H_{\beta}$ lines 
separately for every component. The first spectroscopic orbital solution
was recently derived by \citet{mon07}. They obtained five high-resolution 
echelle spectra, two of which were at the orbital phase $\phi = 0$. 
They also reported a presence of a third component with a velocity 
consistent with the systemic velocity of the binary. The eclipsing pair 
itself was previously identified by the \emph{Hipparcos} satellite 
(\object{HIP 41322}, \object{NLTT 1915}, \object{PPM 715320}) to be a part 
of the visual double system \object{CCDM 08259-1623}. The brightness ratio
from the \citet{mon07} data is consistent with the magnitude difference of $2.37$~mag 
reported in the \emph{Hipparcos} catalogue. Note also that the new 
reduction of \emph{Hipparcos} data \citep{vle07} gives the parallax of 
21.83 $\pm$ 2.25 mas or the distance of 45.8 $\pm$ 4.7 pc.

Eclipses in this system were firstly noted in the ACVS \citep{poj02}. 
For ASAS-08 the dataset spans more than nine and covers the full orbit. 
The ACVS orbital period $P = 1.52852\,d$ is also not perfect.
The ASAS light curve shows no clear evidence of the out-of-eclipse variations. 
However, our data (see Sect. \ref{sect_lc08}) clearly show a presence 
of spots, which are expected in active late-type stars. These spots and the
third light have made our photometric analysis quite challenging.


\section{Observations}

\begin{figure*}
\centering
\includegraphics{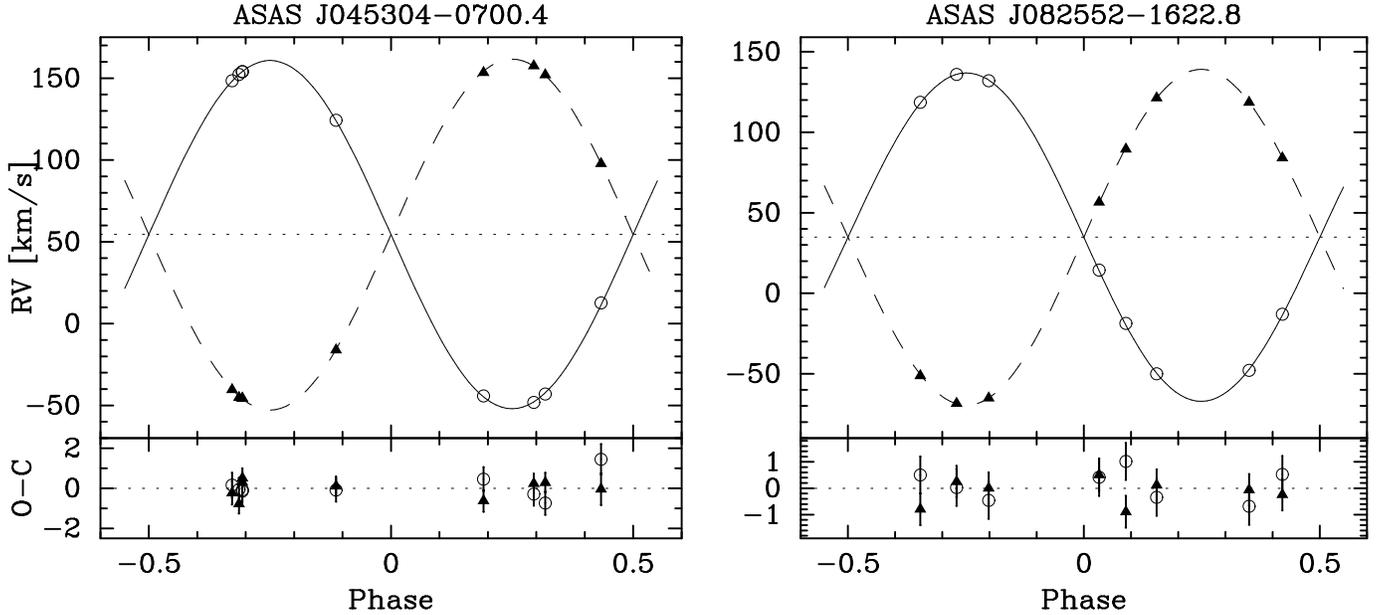}
\caption{Radial velocities of ASAS-04 (left) and ASAS-08 (right) from
	our Keck spectra together with the corresponding best-fitting  
        radial velocity (RV) models. The RVs of the
	primaries are plotted with open circles and those of the secondaries 
	with triangles. The solid line shows the best fit for the primaries 
	and the dashed line that for 
	the secondaries. The dotted line shows the systemic velocity $\gamma$.
	\label{fig_rv}}
\end{figure*}

\subsection{Spectroscopy}
Both ASAS-04 and ASAS-08 are targets from our sample of candidate late-type ASAS 
detached eclipsing binaries (DEBs). In order to derive accurate radial 
velocities (RVs) that would allow us to precisely estimate the masses, 
we obtained a series of high-resolution ($R\sim 60 000$) echelle spectra
with the \emph{HIgh Resolution Echelle Spectrometer} (HIRES) on the 
10-m Keck-I telescope. The spectra were taken during four observing runs 
between September 2004 and October 2005 -- nine observations with the 
exposure times of 900 s for ASAS-04 and eight with the exposure time of 
600 s for ASAS-08. The resulting signal-to-noise ratio for a collapsed 
pixel ($SNR$) was $\sim$110 and 170 at $\lambda \simeq 6000$~{\AA} 
for ASAS-04 and ASAS-08 respectively. The spectra were taken with the 
iodine cell and reduced as described by \citet{kon05}. The 
application of the disentangling technique \citep{kon09} did not 
improve the RVs.

Radial velocities for both systems were obtained with our 
implementation of the two-dimensional cross-correlation technique 
\citep[TODCOR][]{zuc94}. As templates we used synthetic spectra 
computed with the ATLAS code \citep{cas03}. The velocities for every 
epoch are collected in Table \ref{tab_rv}. The formal errors, usually 
between 100 and 200 m s$^{-1}$, were computed from the scatter between
the used echelle orders and are underestimated. To obtain a reduced 
$\chi^2$ equal to $\sim$1, we added in quadrature a constant additional 
error, which for ASAS-04 was 0.61 km s$^{-1}$ for the primary and 
0.44 km s$^{-1}$ for the secondary and for ASAS-08 --- 0.6 and 
0.5 km s$^{-1}$ for the primary and secondary respectively.
These values are nearly equal to the resulting $rms$' of the orbital fit.
Even though the iodine cell is used, the resulting RV precision is hampered
by the rotational broadening of spectral lines in both systems ($v\sin i
\sim 25$ km s$^{-1}$) and the activity of the components, which is 
most likely the origin of the additional RV error. Because of the 
broadening we could not perform any reliable abundance analysis. Many 
weaker lines could not be recognized and the measured equivalent widths or 
depths were too inaccurate\footnote{Our reduced Keck I/HIRES spectra of 
ASAS-04 and ASAS-08 are available upon request. The raw corresponding CCD 
frames are also available in the KOA archive of the Keck I/HIRES spectra.}.

\begin{table}
\caption{Barycentric radial velocities of ASAS-04 and ASAS-08}
\label{tab_rv}
\centering
\begin{tabular}{l r r r r r r}
\hline\hline
Date (TDB) & $v_1$ & $\pm$ & {\it O-C} & $v_2$& $\pm$ & {\it O-C} \\ 
-2450000.0	   & km s$^{-1}$ &  &  & km s$^{-1}$ &  & \\
\hline
{\bf ASAS-04} &  &  &  &  &  & \\
3277.0930 & 148.40 & 0.62 &  0.16 & -40.38 &  0.58 & -0.23\\
3277.1274 & 153.98 & 0.58 & -0.15 & -45.75 &  0.49 &  0.32\\
3329.0271 & 152.32 & 0.58 & -0.06 & -45.08 &  0.49 & -0.78\\
3330.0147 & -48.11 & 0.58 & -0.29 & 157.53 &  0.49 &  0.24\\
3454.7577 & -44.19 & 0.60 &  0.46 & 153.45 &  0.55 & -0.63\\
3456.7734 &  12.64 & 0.77 &  1.44 &  97.76 &  0.80 & -0.04\\
3655.1046 & 154.07 & 0.58 & -0.08 & -45.57 &  0.48 &  0.51\\
3656.1196 & -43.04 & 0.59 & -0.74 & 152.00 &  0.50 &  0.29\\
3657.0408 & 124.30 & 0.58 & -0.09 & -16.05 &  0.48 &  0.11\\
\hline
{\bf ASAS-08} &  &  &  &  &  & \\
3329.1417 & -50.01 & 0.70 & -0.34 & 121.32 &  0.60 &  0.12\\
3330.1258 & 131.99 & 0.70 & -0.46 & -65.06 &  0.60 &  0.00\\
3454.7773 & -47.84 & 0.70 & -0.68 & 118.61 &  0.60 & -0.06\\
3455.8200 &  14.36 & 0.71 &  0.42 &  56.63 &  0.62 &  0.51\\
3456.7692 & 118.61 & 0.70 &  0.51 & -51.23 &  0.61 & -0.79\\
3655.1169 & -13.07 & 0.70 &  0.53 &  84.05 &  0.60 & -0.24\\
3656.1383 & -18.66 & 0.70 &  1.02 &  89.58 &  0.60 & -0.89\\
3657.1184 & 135.97 & 0.70 &  0.03 & -68.36 &  0.60 &  0.25\\
\hline
\end{tabular}
\end{table}

\subsection{Photometry\label{sect_phot}}
The original ASAS light curves of ASAS-04/08 are not sufficiently accurate enough 
for a precise determination of the orbital and physical parameters.
Therefore we carried out a three-week observing run on the 1.0-m Elizabeth 
telescope at the South African Astronomical Observatory (SAAO) in January
2008. We used a $1024 \times 1024$ STE4 camera with the Johnson $V$ and
Cousins $I$ filters. The field of view was $317 \times 317$ arcseconds
(0.31 arcsec/pix). In order to keep the $SNR$ at a relatively high level,
exposure times varied depending on the observing conditions. The data 
were processed with the standard data reduction tasks available in the 
\emph{IRAF} package\footnote{\emph{IRAF} 
is written and supported by the \emph{IRAF} programming group at the National 
Optical Astronomy Observatories (NOAO) in Tucson, AZ. NOAO is operated by the
Association of Universities for Research in Astronomy (AURA), Inc. under cooperative
agreement with the National Science Foundation. http://iraf.noao.edu/}.

For the photometric calibration we selected up to five of the brightest 
stars in the ASAS-04/08 fields and in the fields of our other objects
(about 30 stars in total) and inspected them for variability. 
Instrumental magnitudes were corrected for the atmospheric extinction and 
compared with data from catalogues such
as Tycho-2 \citep{hog00}, USNO-A2.0 \citep{mon98}, 2MASS \citep{cut03}, 
SDSS \citep{aba09}, CMC14 \citep{cmc06} and ASCC-2.5 V3 
\citep[][with updates]{kha01}. Whenever possible, we converted the catalogue 
entries for a given star to the standard Johnson's $V$ band, $B-V$ 
colour or Johnson-Cousins $V-I$ colour. When only $B-V$ was possible, we 
transformed it to $V-I$ with colour-colour relation for dwarfs
given by \citet{cal93}. Average values of $V$ and $V-I$ from three and up to
six different measurements were taken and used to transform our instrumental 
magnitudes to the standard ones. The $rms$ of the transformation was about 0.05 
mag in both bands, which later led to systematic errors in the binaries'
parameters (see next section). Altogether, we
collected about 440 measurements in $V$ and 420 in $I$ for ASAS-04 
and 690 in $V$ and 720 in $I$ for ASAS-08.
The $rms$ of the final fit (see next sections) is 7 mmag in $V$ and 
9 mmag in $I$ for ASAS-04 and for ASAS-08 6 and 8 mmag in $V$ and $I$
respectively. The SAAO light curves are shown in Figs
\ref{fig_lc_04} and \ref{fig_lc_08}. Differential light curves
are also available on-line in the electronic version of the article.


\section{Analysis}

\subsection{Spectroscopic orbits}

\begin{table}
\caption{Orbital parameters of ASAS-04 and ASAS-08}
\label{tab_orb}
\centering
\renewcommand{\footnoterule}{}
\begin{tabular}{l r l r l}
\hline \hline
	&\multicolumn{2}{c}{\bf ASAS-04}&\multicolumn{2}{c}{\bf ASAS-08}\\
Parameter & Value & $\pm$ & Value & $\pm$ \\
\hline
$P$ [d] 		& 1.62221933 & 2.7e-7 	& 1.528488572 & 5.4e-8 \\
$T_0$ [JD-2450000]	& 1871.16139 & 3.4e-4 	& 1869.19935 & 1.4e-4 \\
$K_1$ [km s$^{-1}$] 	& 106.47 & 0.23 	& 101.96 & 0.33 \\
$K_2$ [km s$^{-1}$]  	& 107.21 & 0.20 	& 104.29 & 0.29 \\
$\gamma_1$ [km s$^{-1}$]& 54.54 & 0.21		& 34.94 & 0.25 \\
$\gamma_2-\gamma_1$ [km s$^{-1}$]& -0.33 & 0.29 & -0.26 & 0.33 \\
$q$ 		 	& 0.9931 & 0.0029 	& 0.9777 & 0.0042 \\
$a \sin{i}$ [R$_\odot]$& 6.853 & 0.010 	& 6.233 & 0.013 \\
$e$ 		 	& 0.0 & (fix) 		& 0.005 & 0.002 \\
$\omega [^\circ]$ 	& --- &  ---  		& 84.0  & 6.0 \\
$M_1 \sin^3{i}$ [M$_\odot]$ & 0.8226 & 0.0024  & 0.703 & 0.003 \\
$M_2 \sin^3{i}$ [M$_\odot]$ & 0.8170 & 0.0024  & 0.687 & 0.003 \\
\hline
\end{tabular}
\flushleft
Note: Values of $P$, $T_0$, $e$, and $\omega$ for both systems are 
taken from the combined spectroscopic and photometric analysis performed 
with PHOEBE.
\end{table}

Our first step in the analysis was to obtain a preliminary 
orbital solution with RV measurements independent of the
light curve. We used a simple procedure that fits a double-Keplerian 
orbit and minimizes the $\chi^2$ function with a Levenberg-Marquardt 
algorithm. 

At the beginning we set all parameters free, in particular the orbital 
period $P$, time of minimum (phase = 0) $T_0$, eccentricity $e$, 
argument of periastron $\omega$ and systemic velocities $\gamma_1$ 
and $\gamma_2$ calculated for every component separately. In both binaries
$\gamma_1$ and $\gamma_2$ occurred to be indistinguishable. Hence later we
kept the difference between them fixed to 0. For ASAS-04 we also 
found $e$ to be much less different from 0.0 than its uncertainty,
so we also kept it fixed to 0 later. However, for
ASAS-08 we found $e = 0.005 \pm 0.003$. Hence we left this parameter free
in further analysis, also when light curves were used 
(see Sect. \ref{sect_lc08}).
The error value is only the formal one and it is possible that the 
orbit is exactly circular. Nevertheless, a non-zero
eccentricity orbit better fits the observations than one with $e=0$
and in particular allows one to better reproduce the moments of eclipses,
which seem to be different from 0 and 0.5 phases. For the definition
of the zero-phase -- $T_0$ -- we follow the same convention as in 
\citet{hel09} according to which if $e > 0$, the moment 
$T_0$ does not coincide with any of the eclipses (see: \citealt{hel09} and
references therein).

The final set of spectroscopic orbit parameters is given in Table 
\ref{tab_orb}. Values of the period and zero-phase as well as 
$e$ and $\omega$ for ASAS-08 are from the combined RV + LC analysis.
We did not include the data of \citet{mon07} for ASAS-08. Still, our solution
is quite consistent with theirs. The model RV curves based on the parameters from
Table \ref{tab_orb} are plotted in Fig. \ref{fig_rv} together with 
the measurements. The resulting $rms$ for 
ASAS-04 was 0.60 km s$^{-1}$ for the primary and 0.44 km s$^{-1}$ for the 
secondary and for ASAS-08 -- 0.60 and 0.51 km s$^{-1}$ for the primary and 
secondary respectively. TODCOR allowed us to estimate the flux ratio of
binaries' components in the iodine lines regime, which is similar to the $V$
band. We obtain $F_2/F_1 = 0.72\pm0.12$ for ASAS-04 and $0.75\pm0.11$ for ASAS-08. 

Finally, we verified the quality of the best-fitting orbital RV solution in 
several ways. First, we carried out the tomographic disentangling of the component
spectra and used them with our own procedure of computing RVs 
\citep[for more details see:][]{kon09b}. This way we obtained sets of RVs characterized by 
best-fit $rms$ that was a bit worse (by few 100 m s$^{-1}$) but the resulting best-fitting 
orbital parameters were consistent with those based on TODCOR derived RVs at 
the level of 1 sigma. Second, we carried out a bootstrap (where $\sim$30\% of 
measurements are duplicated by drawing the trail RV data points with 
replacement from the original data sets) estimate of the parameters' 
errors and obtained nearly the same errors as those from the least-squares 
formalism, which indicates a good quality of the orbital solution from 
the $\chi^2$ minimalization.
One should also keep in mind that the orbital RV fit was carried out  
simultaneously for both components with $P$, $T_0$, $e$, $\omega$ and $\gamma$ 
common for both stars, so the effective number of datapoints (measurements) 
is twice the number of observing epochs. Furthermore, to estimate the possible 
influence of systematic errors on the orbital parameters, we carried out a 
Monte-Carlo analysis and found that for any 
derived parameter the systematics are at least one order of magnitude smaller than 
formal uncertainties. Also, maxima in the cross-correlation function were 
well separated, hence we do not expect that our, favored in these cases, 
TODCOR based RVs could possibly lead to an orbital solution that is biased 
in any way as was discussed by \cite{sou07}.

\subsection{ASAS-04's spotted light curve\label{sect_lc04}}

\begin{figure}
\centering
\includegraphics[width=\columnwidth]{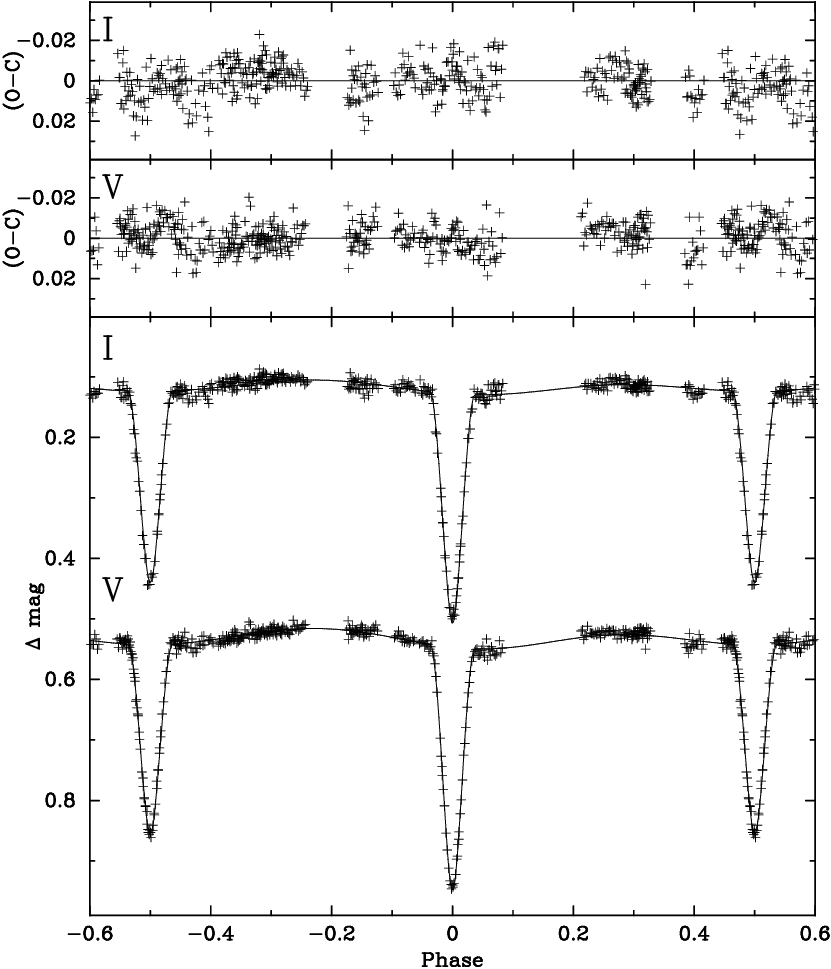}
\caption{ASAS-04 light curve in $V$ and $I$ band phase-folded
	with the best-fitting orbital period $P=1.62221933$ d. The lower panel
	shows (artificially shifted) model light curves together with the 
         data points obtained 
	from the 1.0-m Elizabeth telescope. The two upper panels show the residuals 
	from the best fit. An out-of-eclipse variation is clearly visible.}
         \label{fig_lc_04}
\end{figure}

\begin{table}
\centering
\caption{Absolute physical parameters of ASAS-04\label{tab_par_04}}
\begin{tabular}{l r l r l}
\hline \hline
{\bf ASAS-04} & \multicolumn{2}{c}{Primary} & \multicolumn{2}{c}{Secondary}\\
Parameter & Value & $\pm$ & Value & $\pm$ \\
\hline
inclination $[^\circ]$ &\multicolumn{4}{c}{84.60 $\pm$ 0.03}\\
$a$ [R$_\odot]$  	& 3.430 & 0.008 & 3.454 & 0.008 \\
Mass [M$_\odot]$ 	& 0.8452& 0.0056& 0.8390& 0.0056 \\
Omega-potential 	& 9.130 & 0.025 & 9.202 & 0.026 \\
Radius [R$_\odot]$ 	& 0.848 & 0.005 & 0.833 & 0.005 \\
$\log{g}$ 		& 4.503 & 0.005 & 4.515 & 0.005 \\
$v_{syn}$ [km s$^{-1}$]	& 26.43 & 0.15  & 25.95 & 0.15  \\
$V-I$ [mag] 		& 0.830 & 0.135 & 0.894 & 0.150 \\
	\multicolumn{5}{c}{\it Parameters of spots:}\\
Longitude [rad]		& 3.14  & 0.04  & 2.77  & 0.04 \\
Latitude [rad] 		& 0.85  & 0.06  & 1.0   & 0.1 \\
Radius [rad] 		& 0.71  & 0.02  & 0.72  & 0.02 \\
\% of surface	      & 12.1  & 0.4   & 12.4  & 0.7  \\
Temperat. fraction 	& 0.975 & 0.002 & 0.967 & 0.002 \\
	\multicolumn{5}{c}{\it Including spots:}\\
Effect. temperat. [K]	& 5324  & 200  & 5105 & 200 \\
$M_{bol}$ [mag] 	& 5.46  & 0.16 & 5.68 & 0.17\\
$M_V$ [mag] 		& 5.64  & 0.23 & 5.92 & 0.25 \\
distance [pc]		&\multicolumn{4}{c}{148 $\pm$ 5}\\
	\multicolumn{5}{c}{\it Spots rectified:}\\
Photosph. temperat. [K] & 5340  & 200  & 5125 & 200 \\
Temperature ratio	&\multicolumn{4}{c}{0.960 $\pm$ 0.004$^a$}\\
Luminosity [log(L$_\odot)]$	
			&-0.278 & 0.065&-0.365&0.068 \\
$M_{bol}$ [mag] 	& 5.45  & 0.16 & 5.66 & 0.17 \\
$M_V$ [mag] 		& 5.62  & 0.22 & 5.90 & 0.24 \\
\hline
\end{tabular}
\flushleft
$^a$ The temperature ratio's uncertainty does not include the uncertainty
in the absolute photometric calibration because it is dependent on the eclipse 
depths related only to the relative photometry.
\end{table}

The light curve modelling was done in two steps with two different codes.
The first one -- JKTEBOP \citep{sou04a,sou04b}, based on EBOP 
\citep[\emph{Eclipsing Binary Orbit Program};][]{pop81,etz81}  
-- fits a simple geometric model of a detached eclipsing binary to a single
light curve. The procedure is very fast, and stable and provides an extensive set 
of error-estimate algorithms, but is only suited to relatively well separated 
binaries (such as the two researched here), works with only one light curve at 
a time and does not allow for spots. Nevertheless, with JKTEBOP we derived quite 
accurate values of the parameters, which were later incorporated into the second 
code -- PHOEBE -- to obtain a full model of the system based on the two light 
curves and RVs. We also concluded that for both binaries the preliminary 
results from the two bands treated separately agree well and that the 
analysis performed simultaneously on two light curves is sufficient.

The \emph{PHysics Of Eclipsing BinariEs} \citep[PHOEBE;][]{prs05} code
is an implementation of the Wilson-Devinney (WD) code \citep[with updates]{wil71},
which uses the computed gravitational potential of each star to calculate
the surface gravity and effective temperatures. It also includes theoretical
Kurucz model atmospheres to determine the radiative properties of stellar discs.
It may work with several different light curves and RV curves (one or two) 
at the same time. We used it to obtain the final model of the ASAS-04 system.
The values of some fundamental constants, like the gravitational
constant or the Sun's radius, are slightly different in JKTEBOP and PHOEBE, so 
one has to be careful when incorporating the results from one code into 
the other one.

With PHOEBE, preparing a spotted binary model is relatively easily. 
This software in generally one to add many single spots on both components 
and fit for their size, position and temperature as a fraction of the photospheric 
temperature of the star\footnote{Not to be mistaken with the effective 
temperature. See the explanation for Eq. \ref{eq_teff}.}.

First of all, we improved $P$ and $T_0$ from the spectroscopic solution,
after we had the photometric data and preliminary parameters from JKTEBOP. 
This was possible 
thanks to the PHOEBE's ability of working on several light curves and both RV
curves simultaneously. Later, we kept them fixed and fitted for temperatures, 
gravitational potentials, inclination and quantities obtained from the 
spectroscopic orbit. Spots were added later "manually" by putting their 
parameters and inspecting the light curve by eye. The light curve mimics 
ellipsoidal variations, but a model with two spots, one on each component, 
both seen around eclipses, better fits the observations. 
If the shape of the variation were caused by tidally distorted stars, 
their radii would be much larger and thus eclipses would be much wider. 
Also, a closer inspection reveals an asymmetry between the out-of-eclipse 
variations.

In the last stage of modelling we set all parameters free once again to let 
them converge onto their final values. 
In our analysis we used the square-root limb-darkening law \citep{dia92} 
with the coefficients automatically interpolated by PHEOEBE from the van 
Hamme LD tables \citep{vHa93}. The albedos for both 
components were held fixed at the value of 0.5 as is appropriate for 
convective envelopes. The gravity brightening coefficient was set to 0.32 
-- the classical value obtained by \citet{luc67}.

It is common knowledge that it is hard to reliably estimate temperatures of 
eclipsing binary components. The depth of the eclipse depends not 
only on temperatures, but also inclination, stellar radii and the spots'
configuration, and the ratio of the eclipse depths can give only the ratio 
of effective temperatures. Thus, to compute the stars' temperatures precisely, one 
must have very precise starting values. PHOEBE enables one to separately compute the 
flux from every component in every band. with the $V$ and $I$ light 
curves, we computed the $V-I$ colour for each star and used them with the empirical 
colour-temperature calibration from \citet{cas06}. In this way we obtained a 
starting point for the temperature fitting. This presumably should have 
produced reliable values of the temperatures with relatively small formal 
errors, typically about 50 K. The idea that lays behind this approach is 
described by \citet{prs05,prs06} or \citet{kal09} and is based on the calculation of
the so called \textit{binary effective temperature} -- a time- or phase-dependent
value that can be associated with a colour index of the whole system 
(that is why two or more light curves are required). The discussion however does not 
include the influence of systematic uncertainties (see below) or spots. 
The method itself is known
but was not widely used on spotted systems, which is why more attention at this point 
is required. However, the biggest contribution to the final error budget of 
temperatures in our case comes from the imperfect absolute magnitude 
calibration. The uncertainty of $V-I$, at the level of 0.07 mag, led to
an error of about 200 K in every component's temperature. One should keep in mind that 
this is actually the uncertainty of the absolute scale of temperatures, while 
the ratio is very well constrained. We noticed that a change of any component's 
temperature by 5-10 K results in an eclipse depth variation that is noticeable
with our photometry. We also noticed that temperature variations at a level 
of about 200 K have a small influence on the derived radii and increase their
final uncertainties by about 25 per cent (from 0.004 to 0.005).
Additionally, note that the $V-I$ colour does 
not change significantly even when one uses completely improbable starting 
values of the temperatures for the fitting.

The final set of absolute physical and radiative parameters is listed in 
Table \ref{tab_par_04}. We managed to obtain a very good level of precision
in masses and radii (about 0.6\%). The resulting flux ratio in the $V$ band is 
$0.790\pm0.015$, which agrees with the value from TODCOR, but is 
more accurate. $(V-I)$ values are corrected for the reddening (see below).
Note that the absolute values and 
their uncertainties were not taken from PHOEBE, but were calculated by a 
procedure called JKTABSDIM, which is available together with JKTEBOP.
It provides photometric properties (like absolute magnitudes) and distances
by calculating $M_{bol}$ on the basis of temperature, applying various 
bolometric corrections \citep{bes98,flo96,gir02} to calculate absolute
magnitudes in the required bands and to compare them with the observed values to 
estimate the distance.

Absolute bolometric and $V$ magnitudes were calculated 
with JKTABSDIM in two cases: with and without spots.
In the first case we follow the concept presented by \citet{lop05} and 
calculate a value that we name an \emph{effective temperature} -- $T_{eff}^S$, 
which is actually a temperature related to the true luminosity through the
Stefan-Boltzman law:
\begin{equation}
L \sim \sigma (T_{eff}^S)^4.
\end{equation}
Spotted areas of the stellar surface emit less radiation than 
''clean'' photosphere, thus $T_{eff}^S$ is not equal to the photospheric 
temperature -- $T_{phot}$. The effective temperature is actually the
temperature of a homogeneous, spherical blackbody of the same size 
as the spotted star and radiating the same amount of light as the star with a 
given photospheric temperature and a spot with given parameters. It can be 
calculated from the relation:

\begin{equation}\label{eq_teff}
T_{eff}^S = \left[ 1 - 0.5\, (1-\tau^4)(1-\cos{\theta})\right]^{1/4} T_{phot},
\end{equation}
where $\tau = T_{spot}/T_{phot}$ is a fraction of the temperature of the spot 
surface to the stellar photosphere temperature and $\theta$ is an angular radius
of the spot \citep{lop05}. The spot parameters obtained from our fit are listed in 
Table \ref{tab_par_04}. In PHOEBE the longitude of the spot is counted 
counter-clockwise from 0 to $2\pi$, and 0 means the direction towards the other star. 
The latitude is counted from 0 (the $+z$ pole) to $\pi$ (the $-z$ pole). 

Using $T_{eff}^S$ we can mimic the ''apparent'' properties of the components, 
mainly absolute $V$ magnitudes, and compare them to observed ones. 
Thus we can estimate the distance. We used the parameters from Table 
\ref{tab_par_04} and found the distance of 148~$\pm$~5~pc as a 
weighted average of eight separately calculated results, based on bolometric 
corrections mentioned earlier and in five bands -- our (observed)
$V$ and $I$, and $J$, $H$, and $K_s$ from 2MASS. The best consistency between 
results from various bands was reached for $E(B-V)= 0.13$. The values of
the apparent $M_{bol}$ and $M_V$ are given in Table \ref{tab_par_04} in the 
\emph{''icluding spots''} part.

We ran JKTABSDIM a second time with the photospheric temperatures $T_{phot}$ as 
derived directly by PHOEBE. This way we can calculate the ''true'' values of 
$M_{bol}$ and $M_V$ as if there were no spots at all. They are more 
useful for comparison with theoretical models. We list them in
Table \ref{tab_par_04} in the \emph{''spots rectified''} part. Note 
that the values ''with'' and ''without'' spots are very close. Nevertheless, 
we decided to use the ''rectified'' ones in our further evolutionary status 
analysis and also the age determination, because models are calculated for unspotted 
stars. One should remember that the total large uncertainties 
in colour, temperatures, and magnitudes are caused by the error in the absolute 
magnitude calibration. Formal errors dependent on the quality of the 
photometric data are several times lower. It is particularly important 
for the temperature ratio which is derived with about 0.4\% uncertainty. 

In order to check how the potentials (thus radii) are affected by the calibration
uncertainty, we set the temperatures to be higher by about 200 K, keeping 
their ratio constant. We held the resulting temperatures fixed (at about 
4550 and 4290 K) and repeated the fit.
The resulting potentials in this model have not changed
significantly (well below their uncertainties), therefore we concluded that 
this systematic error in the photometric calibration, colour, and temperatures
does not affect our final results for the radii. We made the same 
conclusion after lowering the temperatures by 200 K from their best-fitting model
values. However, to obtain conservative radii uncertainties, we incorporated the
difference in the radii between those models into the total error of the
radii.

For the spot parameters, the 
temperature fraction and spot's angular radius are generally degenerated
and it is difficult to fit them simultaneously. Some weak constrains are 
possible if the spot is eclipsed -- from the eclipse's shape, and during the 
phases when the spot starts to be visible (larger spots are visible longer).
For of ASAS-04 (and ASAS-08 as well) we are not able to solve that
problem, consequently our solution is degenerated. Errors given in Table \ref{tab_par_04}
are formal only and are probably underestimated. In our analysis, it is 
more important to deduce the influence of the spot on the total flux.
 
Figure \ref{fig_lc_04} depicts our model light curves and SAAO measurements. 
One can easily see the out-of-eclipse variations in the system. 
According to our results, ASAS-04 consists of two slightly inflated 
stars. Theoretical models of the main-sequence stars of these masses predict 
radii around 0.74 R$_\odot$ \citep{bar98}. Also the temperatures exceed 
the values expected for main-sequence dwarfs of given masses by over 
200 K for both components \citep[e.g.][]{har88,tok00} and indicate a 
$\sim$G8 type for the primary and $\sim$K1 for secondary \citep{tok00}. 
This may indicate a main sequence turn-off stage, which is a unique feature 
among known low-mass eclipsing binaries. The quite large difference 
in the eclipses' depths observed in our light curve, especially in the $V$ 
band, implicates the temperatures ratio to be significantly different from 1.
One may find it surprising, considering the nearly equal masses of the ASAS-04 
components. This cannot be fully explained by the age and 
the evolutionary status of the system.

\subsection{ASAS-08: large spots and a third light\label{sect_lc08}}

\begin{figure}
\centering
\includegraphics[width=\columnwidth]{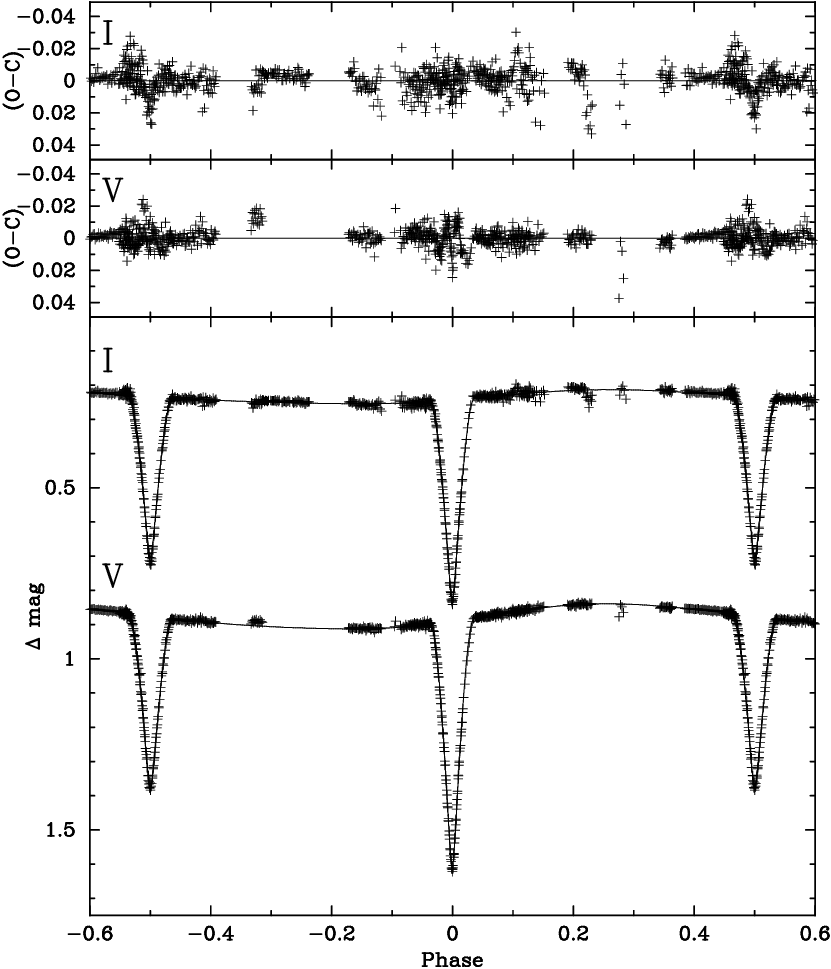}
\caption{Same as Fig. \ref{fig_lc_04} but for ASAS-08 and its best-fit
	orbital period $P=1.52848715\,d$. The variation between eclipses
	comes from two large, cold spots.}
         \label{fig_lc_08}
\end{figure}

\begin{table}
\centering
\caption{Absolute physical parameters of ASAS-08\label{tab_par_08}}
\begin{tabular}{l r l r l}
\hline \hline
{\bf ASAS-08} & \multicolumn{2}{c}{Primary} & \multicolumn{2}{c}{Secondary}\\[2pt]
Parameter & Value & $\pm$ & Value & $\pm$ \\
\hline
inclination $[^\circ]$ &\multicolumn{4}{c}{89.02 $\pm$ 0.13}\\
$a$ [R$_\odot]$  	& 3.082 & 0.010 & 3.152 & 0.009 \\
Mass [M$_\odot]$ 	& 0.703 & 0.003 & 0.687 & 0.003 \\
Omega-potential 	& 9.96 & $_{+0.06}^{-0.12}$ & 9.72 & $_{+0.15}^{-0.12}$ \\
Radius [R$_\odot]$ 	& 0.694 & $^{+0.007}_{-0.011}$ & 0.699 & $^{+0.011}_{-0.014}$ \\
$\log{g}$ 		& 4.600 & 0.016 & 4.585 & 0.016 \\
$v_{syn}$ [km s$^{-1}$]	& 22.9  & 0.4  & 23.1 & 0.4  \\
$V-I$ [mag] 		& 1.26 & 0.11  & 1.40 & 0.11 \\
	\multicolumn{5}{c}{\it Parameters of spots:}\\
Longitude [rad]		& 1.70  & 0.08  & 3.35  & 0.15 \\
Latitude [rad] 	& 0.30  & 0.02  & 1.3   & 0.3 \\
Radius [rad] 		& 1.55  & 0.02  & 0.83  & 0.08 \\
\% of surface		&  49   &  1      & 16    & 3  \\
Temperat. fraction 	& 0.943 & $_{+0.004}^{-0.005}$ & 0.986 & $_{+0.002}^{-0.005}$ \\
	\multicolumn{5}{c}{\it Including spots:}\\
Effect. temperat. [K]	& 4230  & 200  & 4080 & 200 \\
$M_{bol}$ [mag]		& 6.87  & 0.21 & 7.03 & 0.24 \\
$M_V$ [mag] 		& 7.58  & 0.40 & 7.92 & 0.45 \\
distance [pc]		&\multicolumn{4}{c}{44 $\pm$ 9}\\
	\multicolumn{5}{c}{\it Spots rectified:}\\
Photosph. temperat. [K] & 4350  & 200  & 4090 & 200 \\
Temperature ratio	&\multicolumn{4}{c}{0.940 $\pm$ 0.004$^a$}\\
Luminosity [log(L$_\odot)]$	
			&-0.808 & 0.082&-0.909& 0.087 \\
$M_{bol}$ [mag] 	& 6.78  & 0.20 & 7.02 & 0.22 \\
$M_V$ [mag] 		& 7.44  & 0.35 & 7.90 & 0.40 \\
\hline
\end{tabular}
\flushleft
$^a$ The temperature ratio's uncertainty does not include the uncertainty
in the absolute photometric calibration because it is dependent on the eclipse 
depths related only to the relative photometry.
\end{table}

\begin{table}
\centering
\caption{Parameters of the third light in the ASAS-08 solution.\label{tab_3rd}}
\begin{tabular}{l r l}
\hline \hline
Parameter &  Value & $\pm$\\
\hline
\% of flux in V & 10.0 & --- \\
\% of flux in I & 17.0 & 1.2  \\
$V-I$ $[mag]$ & 1.97 & 0.24 \\
$M_V$ $[mag]$ & 9.40 & 0.52 \\
\hline
\end{tabular}
\end{table}

\begin{figure*}
\centering
\includegraphics[width=0.9\textwidth]{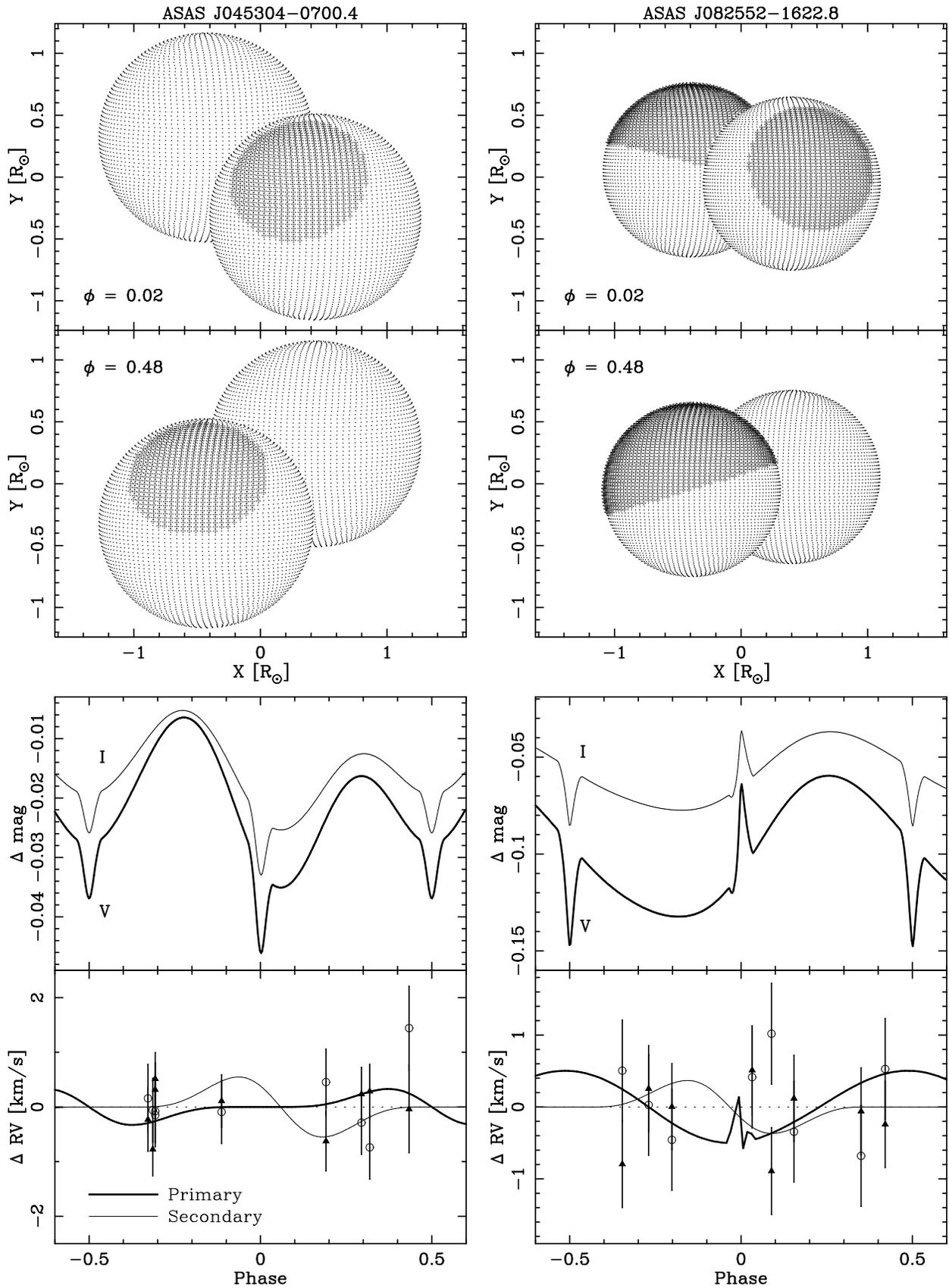}
\caption{Spots on researched systems. 
	{\it Top:} 3D reconstruction of ASAS-04 
	(left) and ASAS-08 (right) during the primary ($\phi=0.02$) and secondary
	eclipse ($\phi=0.48$). Units are solar radii, scale is the same for 
	both systems; {\it bottom:} Influence of spots on the ASAS-04 
	and ASAS-08 light and RV curves. The absolute difference between a spotted 
	and non-spotted case is plotted. Differential RV curves are overplotted 
	on RV residuals, the same as in Fig. \ref{fig_rv}.}
         \label{fig_spot}
\end{figure*}

The analysis of the ASAS-08 light curve was similar to ASAS-04's.
The light curve of ASAS-08 based on our SAAO data is shown in Fig. 
\ref{fig_lc_08}. One can see that the out-of-eclipse variation is clear 
and pronounced. To make the fitting easier and more reliable and the model 
more accurate, we decided to fit a two-spot model with one spot on each component.
One large spot could not explain the modulation.

As previously, we firstly corrected the ephemeris and orbital values by
joining the RVs and two light curves in PHOEBE (the resulting values are given in 
Table \ref{tab_orb}) and then,
keeping them fixed, obtained preliminary values of the temperatures and potentials.
The minimal eccentricity manifests itself in phases of brightnes minima, which
for ASAS-08 are $\phi_{prim} = -0.00018$ or 0.99982 and $\phi_{seco} = 0.50018$.
Spots were added later ''manually'' by putting in their parameters and inspecting
the light curve by eye. The automatic fitting procedure was run only when the 
initial model was reproducing the light curve shapes in a satisfying way. 
From the character of the brightness variation, we deduced that the spots 
should be relatively large and reach the polar regions, at least for one 
component. As for ASAS-04, the solution is degenerated for spot sizes and 
temperature factors.

The last stage was adding the third light into the solution. We aimed to keep
the flux ratio in $V$ unchanged at the level given by the \emph{Hipparcos} 
photometry and \citet{mon07} data (see Sect. \ref{sec_lit}), but we had no
constrains for the $I$ band. Unfortunately, PHOEBE does not allow to
fix the third light in one band and fit for it in the other at the same time.
Hence we decided to find several light curve solutions for a range of the $I$ 
band third light contamination values and compare their quality. Temperatures, 
potentials, inclination and, spot parameters were fitted while the third light 
flux was held fixed. The quality criterion was based on the $\chi^2$ for 
both light curves, shapes of the model curves and the soundness of the obtained 
parameters. 

We found that some parameters, especially the radii, varied
significantly with the third light level in $I$. 
Above a certain value, the resulted $R_1$ started to be smaller than $R_2$,
which is somewhat improbable when the binary is on the main sequence. 
When the third $I$ flux level was too low, we ended up with a 
situation where the third component is hotter (smaller $V-I$) than the binary 
components. If that is the case, and if it is gravitationally bound to ASAS-08, as 
suggested by \citet{mon07}, one should expect it to be brighter in $V$ than 
the binary. We could also reject some solutions by inspections of model light
curves, which for example did not reproduce the depth of one of the minima.
Finally, we found that when we increased the third $I$ flux, 
the $\chi^2_V$ minimum (for the $V$ curve) was getting lower, 
while $\chi^2_I$ (the $I$ curve) was increasing. 
The final level of the third light contamination in the $I$ band -- 
17\% -- was found to be a minimum of the sum $\chi^2_V + \chi^2_I$ 
and occurred to be a compromise between all previously mentioned constrains.
Still the resulting $R_1$ is smaller than $R_2$, but within errors we are 
able to find $R_1 > R_2$. 

Tables \ref{tab_par_08} and \ref{tab_3rd} show the full set of results 
of our modelling. In Table \ref{tab_par_08} we collected stellar 
absolute physical parameters (as in Table \ref{tab_par_04} for ASAS-04). 
The resulting flux ratio in $V$ band is $0.68\pm0.01$, which 
agrees with the value from TODCOR, but is more accurate.
We reached 0.43\% precision in masses and 1 -- 2\% precision in radii.
Most of the errors are caused by the third light $I$ flux's uncertainty.
The formal error of modified potentials is at the level of 0.04
and the rest is the contribution from the third light. As for ASAS-04, 
changing temperatures by 200 K did not affect the potentials significantly. 
Uncertainty in the absolute photometric calibration was however added to 
the colour and the absolute magnitudes error.
Absolute bolometric and $V$ magnitudes were again calculated with 
JKTABSDIM for the two cases with and without spots. In the case with spots, 
we found the distance to ASAS-08 to be 44 $\pm$ 9 pc, which 
agrees very well with the one from \emph{Hipparcos} ($45.8 \pm 4.7$ pc). 
On this basis, we also calculated $M_V$ for the third star. 
Its uncertainty given in Table \ref{tab_3rd} includes the \emph{Hipparcos}
measurement error, which is however not incorporated into the $V-I$
error because the colour is calculated straightforwardly from the model.
The error of the absolute photometric calibration is added to both 
mentioned parameters. We also note that values of $M_{bol}$ and $M_V$ 
for the binary's components are again consistent within error bars but, as for 
ASAS-04, we decided to use the ''rectified'' ones in further analysis.

In order to show the influence of spots in both investigated systems 
quantitatively, in we plot Figure \ref{fig_spot} the absolute difference 
in magnitudes between our final models and models of binaries with the 
same absolute physical parameters but without spots. Additionally we 
plot differential RV curves constructed with PHOEBE as a difference between 
the spotted and unspotted model with the eclipse proximity effects turned
off in both cases so that the Rossiter-McLaughlin effect is not shown.
One can deduce that 
for ASAS-08 a spotted surface is seen all the time because the spot on the
primary covers the polar region. The magnitude of the light variation -- 
$0.03 - 0.05$ -- is not very high. It is in deed lower than the 
peak-to-peak scatter of the ASAS measurements, so from the ASAS 
data only it would be difficult to deduce the presence of spots in our 
systems. Especially if the spots were moving during the nine years of 
observations, which is presumably possible. For this reason
our RV measurements could have been obtained 
when the spot configuration was different. We cannot say much about 
the spot's pattern at the moment of spectroscopic observation. From
the Fig. \ref{fig_spot} we can however estimate that the RV curve's
modulation is smaller than about 0.5 km~s$^{-1}$ for the observed
configuration of spots in our photometric data set.
Obviously, this is within the total RV errorbars.
We also found that turning on the proximity 
effects while fitting to RVs in PHOEBE does not significantly change the results
and in the case of ASAS-08 makes the $rms$ of the RV fit a bit larger. 
We believe the total error of the RV measurements does account
for the overall activity of the targets and therefore the resulting
uncertainties of the best-fitting parameters are realistic.

Figure \ref{fig_spot} also shows three-dimensional 
reconstructions of ASAS-04 and ASAS-08 in two phases -- $\phi=0.02$ 
(primary eclipse) and $\phi=0.48$ (secondary eclipse). Values on axis 
are solar radii and stars are shown on the same scale.

The properties of the spots can be found to be slightly puzzling. Among 
several late-type detached eclipsing binaries known to have spots, 
for example \object{CM~Dra} \citep{lac77,mor09a}, \object{YY~Gem} 
\citep{kro52,tor02}, \object{CU~Cnc} \citep{rib03}, 
\object{GU~Boo} \citep{lop05}, \object{LP~133-373} \citep{vac07} 
or \object{BD~-22~5866} \citep{shk08}, none is reported to be
covered to an extent such large as ASAS-08. Some of seven eclipsing 
binaries listed by \citet{cou07} have more than half of their surface 
covered by spots, but their real masses are only estimated from the 
evolutionary models. In some other cases, like GU~Boo \citep{lop05}, the 
temperature factor is higher than 1 (a ''bright'' spot). It is not 
clear if the existence of such a feature is physically valid, but 
an explanation given by \citet{lop05} suggests that almost whole star's 
surface is covered by many small spots of high contrast and only a small 
fraction of ''clear'' photosphere is seen. This seems to be the case of 
ASAS-08 as well especially having the temperature factors close to 1. 


\section{Age determination and evolutionary status}

\subsection{Activity and kinematics}

\begin{figure*}
\centering
\includegraphics[width=0.9\textwidth]{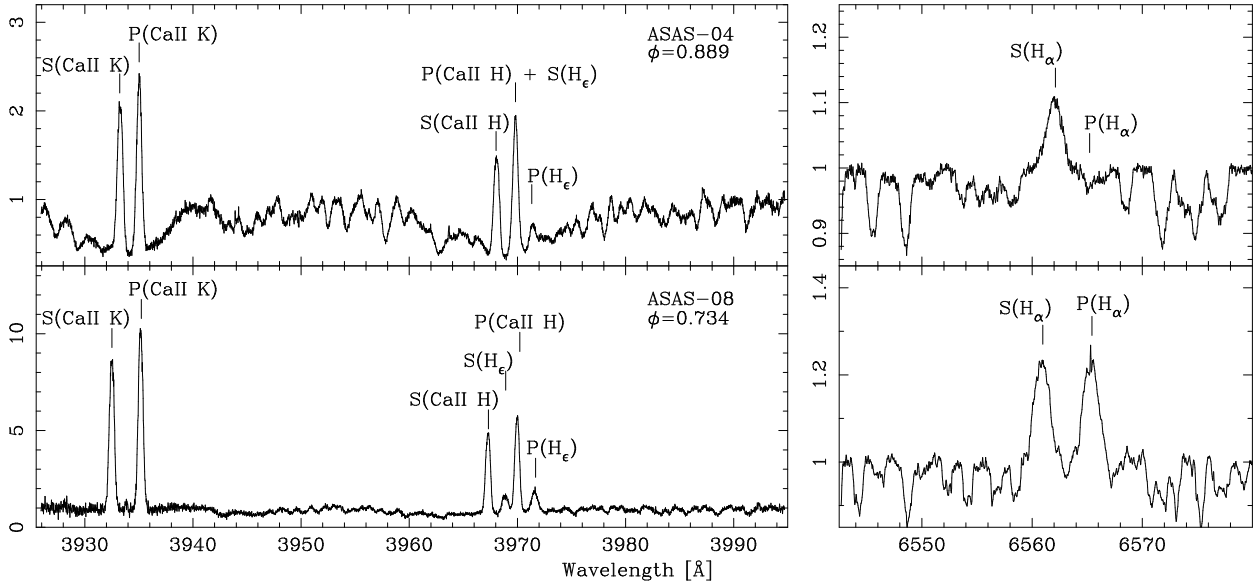}
\caption{Portions of continuum-normalized spectra of ASAS-04 (top) and ASAS-08 
(bottom) around Ca II K and H (left) and H$_\alpha$ (right) lines with marked emission 
features. The primaries features are labelled with P(...) and the secondaries 
with S(...). For ASAS-04 the primary's Ca H line is blended with the secondary's 
H$_\varepsilon$.  Both spectra were taken on October 12, 2005. Orbital phases 
are labelled.}\label{fig_spec}
\end{figure*}

As mentioned above, radii of both ASAS-04 components 
and their higher temperature indicate late main-sequence or even a 
post-main-sequence evolutionary stage.
Inflated radii are commonly observed in low-mass eclipsing binaries,
but together with lower temperatures \citep{tor02,cha07}, which for 
ASAS-04 seem to be higher than predicted. This may still be the effect of 
the absolute magnitude calibration, but it would mean that the real temperatures 
differ by almost 3$\sigma$. Keeping in mind the uncertainty in the temperatures 
one cannot be convinced that ASAS-04 is indeed an old system. But other 
observational facts seem to support this hypothesis.
Young objects of this mass should exhibit quite a high
level of activity. A substantial emission in X-rays and UV should be observed
and a strong emission in spectral lines should also be present. 
No significant UV emission is seen around ASAS-04's position, and
at a distance of 13'' a faint X-ray source can be found -- 
\object{1RXS~J045304.1-070011}. \citet{lop07} found a correlation between
the activity level and radii of low-mass stars in close binaries. The
fractional deviation of a star's radius as compared to 1 Gyr, $Z = 0.02$,
$\alpha = 1.0$ BCAH98 isochrone \citep{bar98} for main-sequence stars 
turned out to be a function of the X-ray-to-bolometric luminosity ratio -- 
$L_X/L_{bol}$. For a linear correlation of $L_X$ and rotational 
velocity \citep[as found by][]{fle89}, and stars $M<0.7$ M$_{\odot}$, 
the relation was
\begin{equation}
\frac{R_{obs}-R_{mod}}{R_{mod}} \propto 74.43 \frac{L_X}{L_{bol}},
\end{equation}
where $R_{obs}$ is the observed radius and $R_{mod}$ is the radius predicted
by the BCAH98 isochrone for a given mass. When stars with $M>0.7$ M$_{\odot}$
were included, the slope changed to 62.511 \citep{lop07}. The uncertainties of 
the slope's values were not given, nor was the zero-point of the relation, but from
Fig. 3. in \citet{lop07} one can deduce that it was about 0.005 and 0.015 for
the cases without and with the more massive stars respectively.

For the ASAS-04 we found $(L_X/L_{bol})_1 = 7.0\pm2.5 \cdot 10^{-4}$ and 
$(L_X/L_{bol})_2 = 8.4\pm3.0 \cdot 10^{-4}$ which corresponds  
to 5 and 6\% excess of the radii on the main sequence respectively. Our measurements 
indicate $13.0\pm0.7$\% for the primary and $11.7\pm0.6$\% for the secondary, 
which disagrees with the relation even when the large scatter of the \citet{lop07}
data points is taken into account.

In the spectra of ASAS-04 we detected a weak emission in H$_\alpha$ for 
the secondary component and for the primary the absorption line is almost 
filled-in. In terms of H$_\alpha$ equivalent width (see below) we
can consider ASAS-04 as ''weakly active''. However,
we also detected a substantial emission in the Ca II H and K lines, as well as 
H$_\varepsilon$. Balmer line H$_\beta$ was in the region between ''blue''
and ''green'' chips of the HIRES detector, so it could not be subtracted. Marks of 
emission in H$_\delta$ and H$_\gamma$ can also be found, but blending with other 
lines made them too difficult to measure. No emission was found in helium lines 
(4027, 4473, 5877 and 6678 \AA) or in the Na I D$_1$ and D$_2$ lines, which suggests 
no variability in the lower chromosphere \citep{mon97}.
We list the equivalent widths given in 
\AA for all emission lines measured in Table \ref{tab_eqw}. 
Portions of spectra around H$_\alpha$ and Ca II lines are depicted in Fig. 
\ref{fig_spec}. The clear existence of some activity indicators is 
nevertheless not surprising. Recent results for low-mass detached eclipsing 
binaries suggest that the stellar activity is excited by the presence of
a close companion \citep[see e.g.][]{lop07}, and the presence of spots
is quite common among those objects. A recent spectroscopic research by 
\citet{par09} revealed that the $H_{\alpha}$ emission is probably variable.
Their low-resolution spectra show only filled-in lines and no emission components
at the level similar to the one presented here.

\begin{table}
\centering
\caption{Equivalent widths of emission lines of ASAS-04 and ASAS-08\label{tab_eqw}}
\begin{tabular}{cccccc}
\hline
 & &\multicolumn{2}{c}{\bf ASAS-04} &\multicolumn{2}{c}{\bf ASAS-08}\\ 
Line & $\lambda$ [\AA] & Pri. & Sec. & Pri. & Sec. \\
\hline \hline
H${_\alpha}$ & 6563 & 0.19 & -0.03 & 0.65 & 0.55 \\
H${_\varepsilon}$ & 3970 & 0.34 & 0.36 & 0.92 & 1.05 \\
Ca II H & 3968 & 2.02 & 2.12 & 3.17 & 3.66\\
Ca II K & 3934 & 2.66 & 2.79 & 5.31 & 5.55\\
H${_6}$ & 3889 & --- & --- & 0.4 & 0.4\\
\hline
\end{tabular}
\end{table}

In order find additional clues that point to ASAS-04 as an old object, we also 
calculated the galactic space velocities $U,V,W$\footnote{Positive 
values of $U$, $V$, and $W$ indicate velocities towards the Galactic center, 
the direction of rotation and the north pole respectively.} with regard to the 
local standard of rest \citep[LSR;][]{joh87}. We applied our values of the
radial systemic velocity and distance estimate together with the
proper motion of $\mu_{\alpha}=12.02$ mas/yr and 
$\mu_{\delta}=-46.12$ mas/yr from the PPMX catalogue \citep{ros08}. 
Values of $U=-15.6 \pm 0.9$, $V = -43.1 \pm 1.4$ and 
$W = -26.1 \pm 1.1$ km s$^{-1}$ put ASAS-04 outside of any known young moving 
group or group candidate \citep{zha09}, being closest to the Hercules stream
\citep{sea07}, and at the transition area between the thin and thick galactic 
disc \citep{ben03,nor04}. \citet{ben07} investigated the age and abundance
of the Hercules stream and found that this feature is composed of a mixture 
of young and old stars, all of which show kinematic characteristics similar to the
thick disc. Thus we may conclude that ASAS-04 is an old system, presumably 
from the thick disc. The maximal age of the thin disc is not well constrained,
but all researches agree with \citet{jim98} who derived $t_{TD}>8$ Gyr.
Although this value seems to be secure for the lower limit of the ASAS-04 age, 
considering other facts (see next section) we adopt 5 Gyr as the lower limit.
If ASAS-04 is older than 8 Gyr it would mean that it is currently close to the 
turn-off point and ends its main-sequence evolution, which makes it a very 
interesting object for more research.

On the contrary, ASAS-08 is a younger, likely a main-sequence binary. 
With our value of the systemic radial velocity, the \emph{Hipparcos} 
distance, and the PPMX proper motion ($\mu_{\alpha}=-183.7$ mas/yr and 
$\mu_{\delta}=15.49$ mas/yr) we obtained $U=-26.9 \pm 0.8$, 
$V = -20.6 \pm 0.3$ and $W = -18.0 \pm 1.2$ km s$^{-1}$. Those values 
put ASAS-08 exactly in the thin disc regime, but it is hard to associate 
the binary with any young moving group, so we can conclude that its age 
is more than $\sim 1$~Gyr. The level of activity supports this conclusion. 
\citet{haw99} have proven that
M dwarfs in young open clusters have a well defined $V-I$ colour at which their 
activity, measured from the H$_\alpha$ emission, becomes ubiquitous. 
Active stars (which have EW$_{H_\alpha}$ $\ge 1$ \AA) from a certain cluster 
are redder than this characteristic $V-I$. \citet{haw99} gave a relation
between the age $t_{H_\alpha}$ of the cluster and this ''H$_\alpha$-limit'' 
colour at which M dwarfs have EW$_{H_\alpha} \simeq 1$ \AA:

\begin{equation}\label{eq_vi}
V - I = -6.91 + 1.05(\log{t_{H_\alpha}}) .
\end{equation}
Later \citet{giz02} also used this relation for field M dwarfs. Both ASAS-08 
components have EW$_{H_\alpha}$ $\simeq 0.6$ \AA \citep[which agrees with ][]{chr02}, 
consequently we can estimate the lower limit for the system's age with Eq. \ref{eq_vi}.
With our value of the secondary's $V-I$ we conclude that the system must be much 
older than 85 Myr. If we further assume that the third component is inactive, in the
meaning of EW$_{H_\alpha}$, we get a lower limit for the age of the system to be 
about 290 Myr. As for ASAS-04, according to the definition by \citet{haw99},
we consider ASAS-08, as well as ASAS-04, as ''weakly active'' in terms of 
H$_\alpha$ equivalent width (and in those terms only).
ASAS-08 was also observed by \citet{par09}, who recorded substantial emission in
the H$_\alpha$ line at a level similar to the one presented here.

Other activity indicators are also detected in emission, especially the calcium 
H and K lines are very strong (see Fig. \ref{fig_spec} and Table \ref{tab_eqw}). 
As for ASAS-04, other Balmer lines are also seen in emission, including weak 
H$_6$ (out of the range of Fig. \ref{fig_spec}). Other lines,
such as sodium D or He~I 5877 \AA, were not found in emission, 
but \citet{par09} noted them to have ''comparable levels of emission'', which
presumably may indicate their variability. This implicates the 
level of activity is higher than for ASAS-04 and is consistent with the presence of 
larger spots. Also the X-ray-to-bolometric luminosity ratio is higher 
than for the former system. When we incorporated the data for  
\object{1RXS~J082551.4-162244} from ROSAT \citep{vog99} and assumed a 
linear correlation between $L_X$ and $v_{rot}$, we found 
$(L_X/L_{bol})_1 = 7.6\pm3.5 \cdot 10^{-4}$ and 
$(L_X/L_{bol})_2 = 9.5\pm4.0 \cdot 10^{-4}$ for the primary and secondary 
respectively. The excess of the radii, measured in the same manner as 
previously, is $8^{+1}_{-2}$\% and $10.5^{+2.0}_{-1.5}$\% respectively.
Those numbers agree within errors with the relation found by \citet{lop07},
but place ASAS-08 above it.

\subsection{Isochrones}

In this section we compare our results with several popular and widely used sets of
theoretical evolutionary models and isochrones --- Y$^2$ 
\citep{yi01,dem04}, BCAH98 \citep{bar98}, PADOVA \citep{gir00,mar08} and GENEVA
\citep{lej01}. 

\subsubsection{Inflated ASAS-04}

\begin{figure*}
\centering
\includegraphics[width=0.8\textwidth]{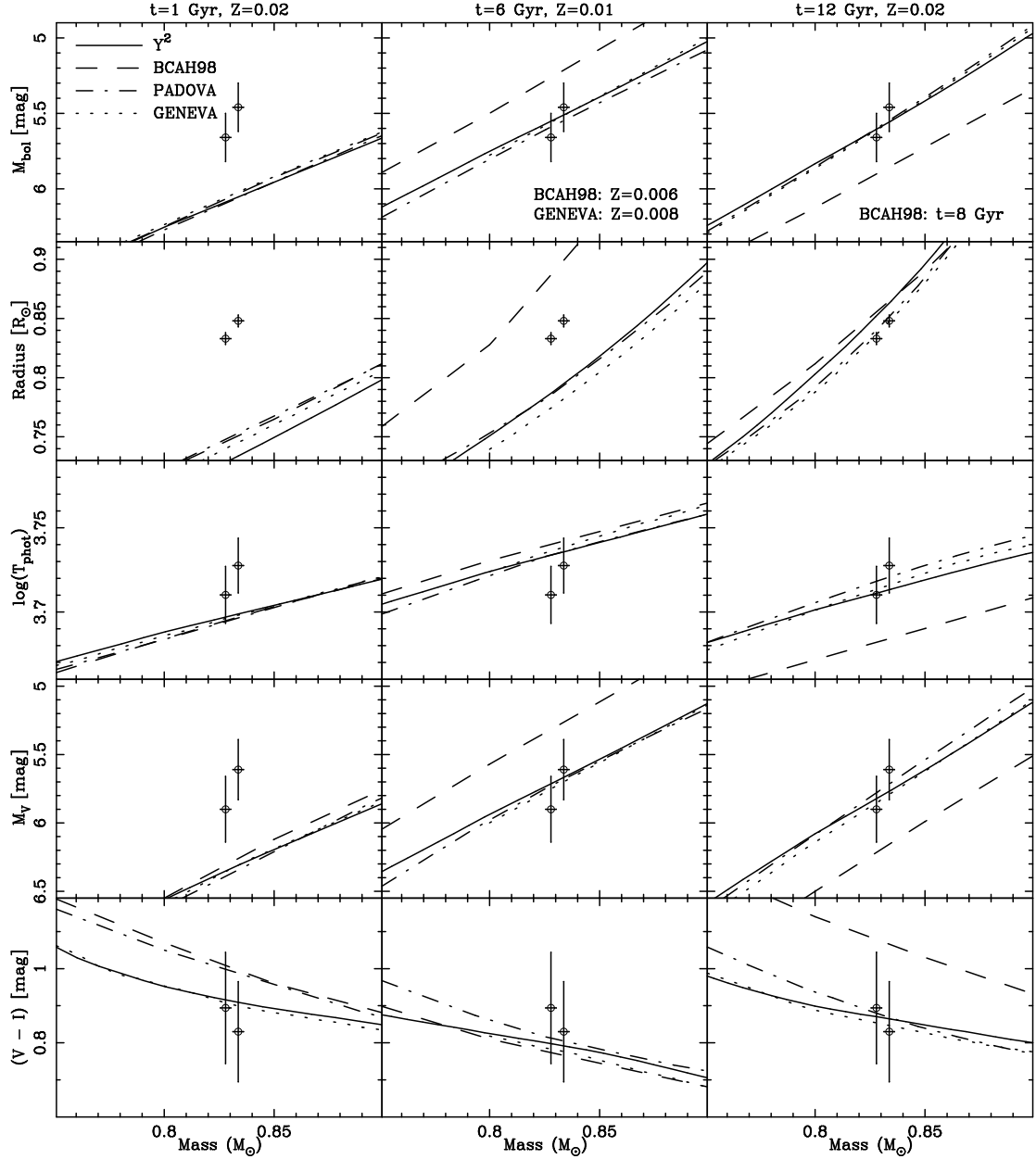}
\caption{Comparison of several physical parameters as a function of mass for 
	ASAS-04 with several sets of isochrones. The left column is for the age of 
	1 Gyr and solar metallicity, the middle column is for 6 Gyr and metallicity below
	solar (0.01 for PADOVA and Y$^2$, 0.008 for GENEVA and 0.006 for BCAH98),
	the right column is for solar composition and $t$ = 12 Gyr, except BCAH98 (8 Gyr).
	From the highest to the lowest, the rows show mass vs. $M_{bol}$, $R$, 
	$\log(T_{phot})$, $M_V$ and $(V-I)$ colour. The measurements and
        their uncertainties are taken from Table \ref{tab_par_04}.}\label{fig_evo_04}
\end{figure*}

In Fig. \ref{fig_evo_04} we show the results for ASAS-04 together 
with several isochrones on the mass versus bolometric magnitude 
$M_{bol}$, radius $R$, logarithm of temperature $\log(T_{phot})$, absolute 
$V$ magnitude $M_V$ and observed $(V-I)$ colour planes. We compare our results
with theoretical predictions for three scenarios: (1) the early main-sequence
stage and solar composition (\emph{left}); (2) late main-sequence stage
and reduced metallicity (\emph{middle}); (3) main sequence turn-off 
and solar metallicity (\emph{right}). Because of the limited availability 
our sets are not uniform for the scenario (2) and (3) -- slightly different 
metallicities were used for the GENEVA and BCAH98 sets in (2) and a different
age for BCAH98 in the scenario (3). In this case we may consider this isochrone
as a lower limit for the age estimate as it usually lies close to one of 
the errorbars. Isochrones for the scenarios (2) and (3) were selected to fit our
measurements in the $M/M_{bol}$ plane. We chose this plane as a reference
since recent findings show that although radii and temperatures are not
particularly well reproduced, the luminosity (hence $M_{bol}$) is and one 
can then fit an isochrone to the data if the radius is scaled by a factor 
$\beta$ and the temperature by $\beta^{-0.5}$ \citep{cha07,mor09b}.

The left column clearly shows that ASAS-04 cannot be at the 
early main sequence stage. The observed properties are not reproduced in 
any of the panels, except for $M/(V-I)$. As was mentioned above, 
both the radii \emph{and} temperatures exceed predictions. 
One could fit a 1 Gyr isochrone to our data if $Z<0.006$
was assumed. However, considering the activity-radius relation \citep{lop07}
and kinematics, we find this scenario very unlikely.

The degeneration between age and metallicity of models that fit the data is
already well known \citep{las02} and ASAS-04 is no exception. Our 
measurements of mass and $M_{bol}$ are well reproduced not only by models 
from Fig. \ref{fig_evo_04}, but also for example for $t = 8$ Gyr, 
$Z = 0.012$. One can also see how BCAH98 for $t=6$ Gyr and $Z=0.006$ varies
from the other isochrones. For the presented case of $t=6$ Gyr, we observe the 
aforementioned discrepancies in the radii and temperatures (except for 
BCAH98, which is for smaller $Z$). Thus we find this age/metallicity 
set a probable one. We observed the same behaviour but with smaller discrepancies
for sets with $t = 8$ Gyr and $Z=0.012$. For the case of 6 Gyr and $Z=0.01$ a 
single correction factor $\beta\equiv R_{obs}/R_{mod}$ can be found for both 
components and would be $\beta\simeq 1.055$. This corresponds to the shift in
$\Delta\log(T_{phot}) \simeq 0.012$, which would allow for fitting a single 
isochrone. The discrepancies diminish around $t \sim 10-11$ Gyr, 
where the radii and temperatures, as well as $M_V$ and $(V-I)$ are nicely 
reproduced. For 12 Gyr the predicted radii start to be larger than observed, 
while in the other planes we obtain a good fit. Thus we consider 12 Gyr as an upper
limit to the ASAS-04 age. Note also that the
BHAC98 model for 8 Gyr and the solar $Z$ predicts the largest radii of all sets, 
despite predicting the lowest temperatures and luminosities. 

The presented models suggest that the radius and temperature discrepancies 
may not be significant for older stars. This seems to be supported by the 
recent discovery of a 0.88 + 0.86~M$_\odot$ evolved eclipsing binary 
\object{V69-47Tuc} in the famous globular cluster \citep{tho09}.
Several sets of theoretical models succeeded to fit the observed radii and bolometric 
luminosities of this binary components with a single isochrone. 
The estimated age was 11.3 Gyr and 
[Fe/H]=-0.70 was assumed. Considering the similar masses of the ASAS-04 components we 
may expect that the almost perfect fits of $t>10$ Gyr isochrones are plausible.

From the discussion above, we can deduce the age of ASAS-04 to be $5 - 12$ 
Gyr and the metal abundance between 0.008 and 0.02 with the ranges of $8 - 11$ 
Gyr and $Z$ from 0.012 to $\sim0.018$ be the most probable ones. This makes it 
one of the oldest low-mass eclipsing systems known to date. Recently several 
evolved low-mas eclipsing binaries were found in 
globular clusters: a 0.81 + 0.68 M$_\odot$, older than 10 Gyr OGLEGC~17 in 
\object{$\omega$ Cen} \citep{tho01}, 0.945+0.144 M$_\odot$ system 
V209 in the same cluster\footnote{This system probably went through a 
common-envolope phase.} \citep{kal07}, a 12 Gyr-old, 
0.79 + 0.23 M$_\odot$ binary V32 in \object{NGC~6397} \citep{kal08},
and the already mentioned V69-47Tuc \citep{tho09}.
In all these cases the primary begins to evolve onto the sub-giant branch. 
Also a few slightly more massive systems were found to be about 10-11 Gyr old.
These are:
\object{CG Cyg} \citep[0.94 + 0.81M$_\odot$; ][]{pop94},
\object{RW Lac} \citep[0.93 + 0.87M$_\odot$; ][]{lac05} and 
\object{HS Aur} \citep[0.90 + 0.88M$_\odot$; ][]{pop86}.
ASAS-04 appears to be similar but less massive than HS~Aur. Hence it
appears to have the smallest masses for a rare group of late-type 
field eclipsing systems that come to the end of their main-sequence 
life.

One should keep in mind that the isochrones fit to the measurements thanks to 
the relatively large errorbars caused by the systematic error contribution of the
absolute photometric calibration. Still, the temperature ratio and therefore the 
magnitude difference are constrained well and none of the models predict 
them correctly.

\subsubsection{Main sequence ASAS-08}

\begin{figure*}
\centering
\includegraphics[width=0.8\textwidth]{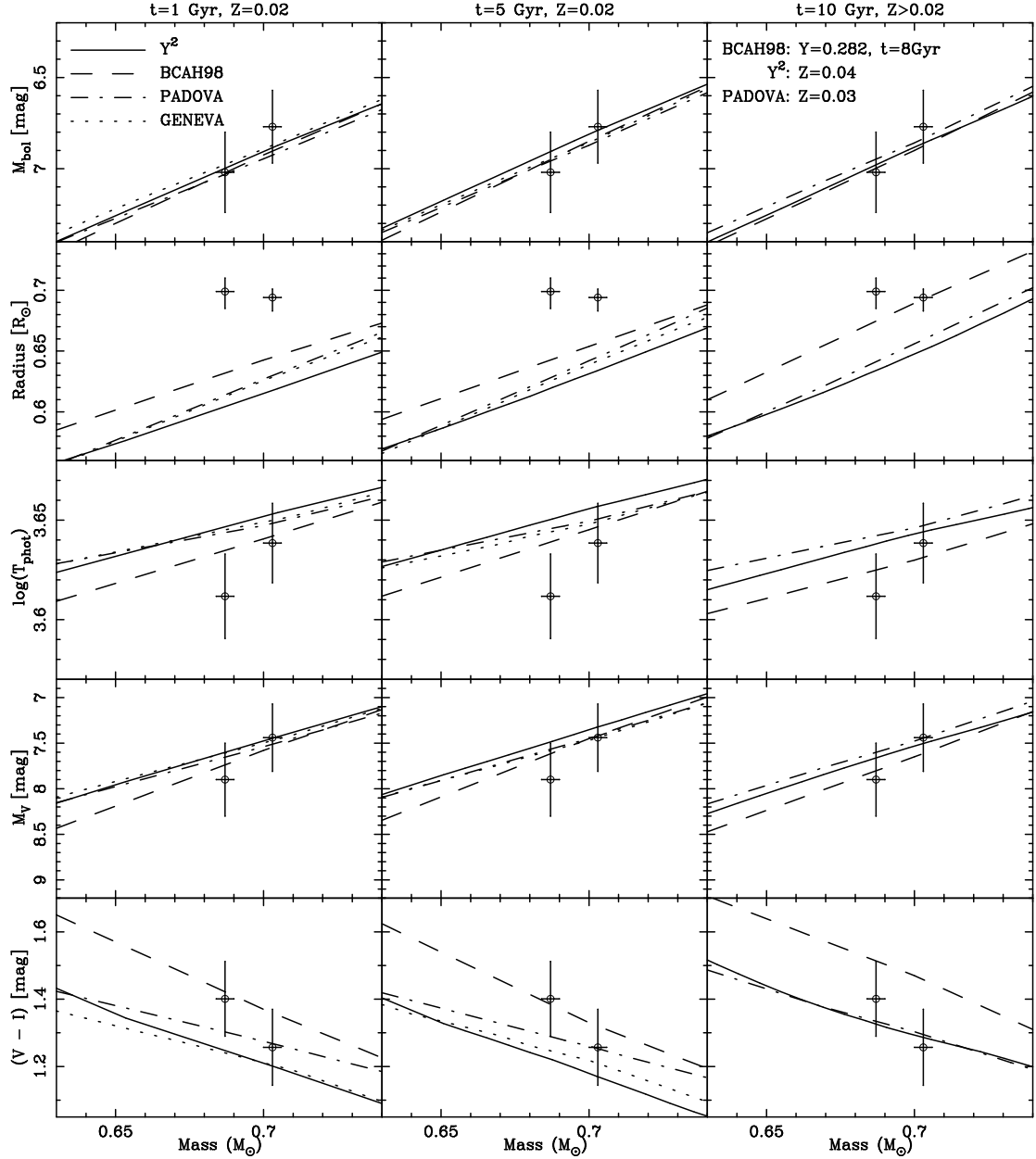}
\caption{Same as Fig. \ref{fig_evo_04} but for ASAS-08. Different ages and 
	metallicities were used: $Z=0.02$ with $t=1$ (left) and 5 Gyr (middle),
	and $Z>0.02$ for $t=10$ Gyr (right column). For the last case the GENEVA set was not 
	available for masses below 0.8 M$_\odot$ and other sets' metallicities 
	vary, as labeled. As for ASAS-04, BCAH98 set was available only for 8 Gyr
	and a helium abundance value $Y=0.282$.}
	\label{fig_evo_08}
\end{figure*}

We show a comparison of our results for ASAS-08 with the isochrones in Fig. 
\ref{fig_evo_08}. This time we selected isochrones for ages $t = 1$ and 5.0 Gyr 
with the solar abundance (left and middle) and 10.0 Gyr and Z higher than solar. 
Because of the limitations of the available models, our 
sample is not uniform. The GENEVA set of models lacks isochrones for masses 
below 0.8 M$_\odot$ and $Z>0.02$, that is why they are plotted only in 
two columns. Other sets are also limited in high metallicities: PADOVA has its 
maximal $Z=0.03$, Y$^2$ are plotted for $Z=0.04$. The BCAH98 model in the 
right column is prepared for the helium abundance of $Y=0.282$ and maximal 
age of 8 Gyr. An isochrone with this $Y$ reproduces properties of the Sun
at its current age \citep{bar98} and for $t=8$ Gyr also fits quite nicely 
to ASAS-08.

As for the previous system, we select models that fit our data in the 
mass-bolometric magnitude plane. ASAS-08 is again a case of the degeneration
between age and metallicity, but we may estimate a lower limit of $Z$
to be about 0.012. Not much difference is seen in the main-sequence 
stage (between 1 and 5 Gyr in Fig. \ref{fig_evo_08}), but to fit an isochrone 
with a larger $Z$, one needs to go to 8 -- 10 Gyr regime. For such a high age 
theoretical models seem to match our data quite well, but this scenario is 
unlikely when considering activity and kinematics. Thus a main sequence 
evolutionary stage of ASAS-08 is very probable.

As for most of other known low-mass eclipsing binaries, for ASAS-08 we observe
clear discrepancies between the models and observations in radii and temperatures
on the main sequence. Every model we compare requires a slightly different 
correction factor $\beta\equiv R_{obs}/R_{mod}$ and within errorbars one can 
find a single value of $\beta\simeq1.09$ for both components. This corresponds 
to a shift of 0.02 in $\log(T_{phot})$, which allows for fitting a single isochrone.

\begin{figure}
\centering
\includegraphics[width=\columnwidth]{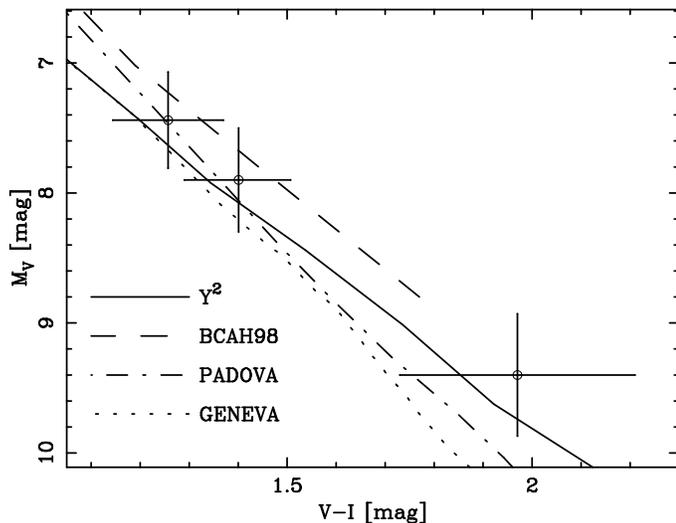}
\caption{$(V-I)/M_V$ plane with the theoretical isochrones of 1 Gyr and $Z=0.02$, 
	and our results for the three bodies in the ASAS-08 system. Data points and 
	uncertainties are from Tables \ref{tab_par_08} and \ref{tab_3rd}. The BCAH98 model 
	ends for 0.6 M$_\odot$.}
	\label{fig_3rdlight}
\end{figure}

One can notice that around $M \sim 0.7$ M$_\odot$ various sets predict 
significantly different $(V-I)$ colour, which is of course related to the different 
physics used in each model. In our case only the PADOVA and BCAH98 sets 
succeed to match the data within errorbars. The BCAH98 set fits slightly better for 
5 Gyr while PADOVA for 1 Gyr. With those two sets and our estimates of 
$(V-I)$ and $M_V$ for the third body, we can also estimate the mass of this star. 
In Fig. \ref{fig_3rdlight} we show our estimates of $(V-I)$ and $M_V$ for 
all three bodies in the ASAS-08 system. For the comparison we plot all sets, and all fit 
within errorbars but only PADOVA and BCAH98 fit simultaneously to the mass measurements.
PADOVA (dot-dashed) gives the consistent third body mass estimate for 
$M_3 = 0.43 - 0.5$ M$_\odot$ and BCAH98 (dashed) ends at $0.6$ M$_\odot$, which
in the colour/brightness plane is close to the edge of errorbars.
Thus $0.6$ M$_\odot$ is an upper limit of the third body's mass.
These values put the third component into the late K spectral type range 
\citep[$T_{eff}\sim4000$~K as predicted by][]{har88}. The $V-I$ colour itself
predicts a fairly later type around M2 \citep[$T_{eff}\sim3600$~K, based on][]{bes98}.
The angular distance is 0.845'' (from \emph{Hipparcos}), which corresponds to
a projected separation of 39 AU if gravitationally bound. We expect that 
a $\sim 0.5$ M$_\odot$ star at this distance would cause a significant 
perturbations in the binary, which presumably could be detected with a 
long-term RV or eclipse timing.


\section{Discussion and conclusions}

We presented an analysis of two new eclipsing systems with component 
masses below 1 M$_\odot$ -- \object{ASAS J045304-0700.4} and 
\object{ASAS J082552-1622.8} -- the first two of our sample of low-mass 
eclipsing binaries from the ASAS database. Both objects seem 
to be interesting targets for further studies: ASAS-04 because of its 
evolutionary status possibly at the end of the main-sequence 
stage and ASAS-08 because of the activity, the unusually large spot coverage and 
the third component. The third body may be responsible for the low
but measurable eccentricity of this system's orbit.

\begin{figure}
\centering
\includegraphics[width=\columnwidth]{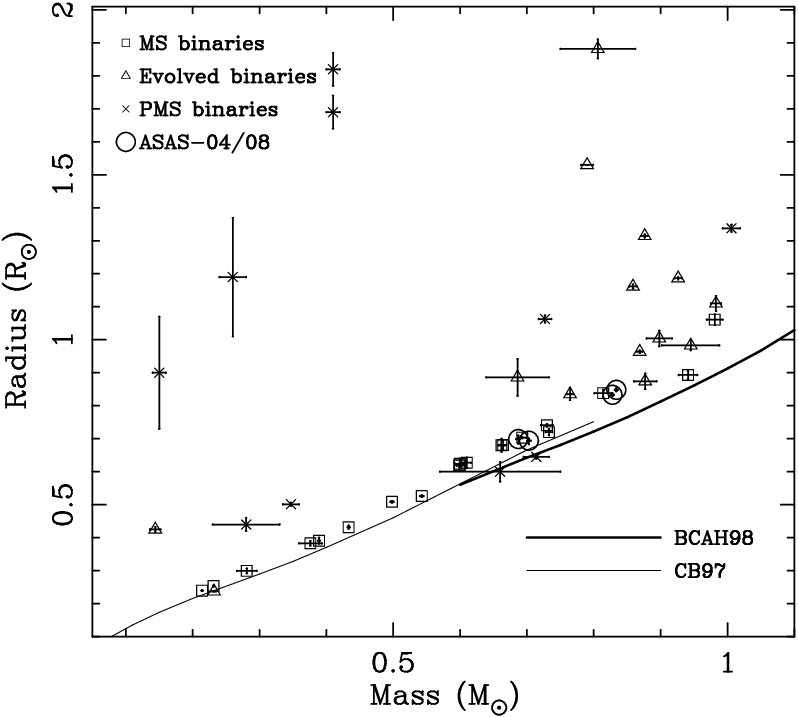}
\caption{Mass/radius data for eclipsing binaries with masses below 1~M$_\odot$. 
	For main-ssequence binaries only those with $M$ and $R$ uncertainties below 3\%
	were chosen from \citet{tor02,rib03,shk08,irw09,mor09a,mor09b,roz09,tor09}. 
	Data for evolved ($t\gtrsim 10$ Gyr) systems were taken from: 
	\citet{pop86,pop94,kal07,kal08,tho01,tho09,lac05}. Pre-main-sequence systems 
	data were taken from: \citet{irw07,sta07,sta08,cak09b,cak09a}. The BCAH98 model 
	for $t$ = 1 Gyr and $Z=0.02$, which ends for 0.6 M$_\odot$ (thick solid line), 
	is suplemented by a CB97 isochrone, available for masses below 0.8 M$_\odot$ 
	(thin line).}
	\label{fig_all_MR}
\end{figure}

Our mass and radius measurements (open circles) are compared with the
other known low-mass eclipsing binaries in Fig. \ref{fig_all_MR}. 
We selected main-sequence systems 
with parameters derived with an accuracy better than 3\% (open squares), 
as well as several systems with ages higher than about 10 Gyr (open triangles) 
and systems with pre-main-sequence components (crosses). A BCAH98 
isochrone for $t=1$ Gyr, $Z=0.02$ and $M>0.6$ M$_\odot$ is also shown, 
supplemented by a similar isochrone from \citet[][CB97]{cha97} set,
for masses below 0.8 M$_\odot$\footnote{We did not use this set in our
analysis because its mass range does not reach ASAS-04 values.}.

We derived both objects' absolute physical parameters with a precision of 
at least single \%. However, uncertainties of some of the values like the
temperatures can likely still be improved because their systematic errors 
are quite large. 
PHOEBE's way of calculating the temperatures based on the passband luminosities 
and colour-temperature calibrations \citep[more details in][]{prs06} 
produces low formal errors, but an uncertainty coming from the photometric 
calibration also contributes. The absolute scale of the temperatures is uncertain, but 
their ratio, highly dependent on the eclipse depths, is constrained rather well. Its 
value for ASAS-04 seems to be too far from unity for stars of these masses. 
There seems to be a discrepancy between the theoretical and observed radii and 
temperatures for ASAS-08 and ASAS-04 if they are on the main sequence, 
which is common for low-mass binaries with main
sequence components (see Fig. \ref{fig_all_MR}). But if ASAS-04 is old,  
the isochrones of about 10 -- 11 Gyr match the measured radii, temperatures,
and luminosities simultaneously.

Our orbital/physical models, which predict large areas of both stars in both systems 
to be covered by spots are not the only possible solution to explain the 
light curve's shape. However, in the case of a different spot scenario, the 
component masses and radii should not change significantly. Nevertheless, we 
tried to obtain spot models as robust as possible, and the large difference in the
temperatures of the ASAS-04 components appears to be real and challenges
the evolutionary models. The isochrones in general succeed to reproduce 
the results for both systems (with the exception of ASAS-08 radii and temperatures,
as expected for the main sequence, and temperature ratio in ASAS-04) and our 
distance estimation for ASAS-08 based on the ''effective'' 
temperatures agrees very well with the \emph{Hipparcos} value. This 
supports the statement that our orbital/physical models are realistic
and accurate.

\begin{acknowledgements}
We would like to thank Grzegorz Pojma\'nski for pointing out several suitable
ASAS targets in the early stages of this project.
KGH would like to thank John Menzies from the South African Astronomical 
Observatory for his support during observations, Janusz Ka\l u\.zny from 
the NCAC Warsaw for useful advices and Tomasz Kami\'nski from the NCAC Toru\'n 
for his help with spectra reduction. 

The authors wish to recognize and acknowledge the very
significant cultural role and reverence that the summit of
Mauna Kea has always had within the indigenous Hawaiian
community. We are most fortunate to have the opportunity
to conduct observations from this mountain.

This work is supported by the Foundation for Polish
Science through a FOCUS grant and fellowship, by the
Polish Ministry of Science and Higher Education through
grants N203 005 32/0449 and 1P03D-021-29. 

This research was co-financed by the European Social Fund and the national
budget of the Republic of Poland within the framework of the Integrated
Regional Operational Programme, Measure 2.6. Regional innovation
strategies and transfer of knowledge - an individual project of the
Kuyavian-Pomeranian Voivodship "Scholarships for Ph.D. students
2008/2009 - IROP"

\end{acknowledgements}

\bibliographystyle{aa}

\end{document}